\documentclass[preprint,aps]{revtex4}

\usepackage{graphicx}

\begin{document}
 \title{Approximate pre-classical solutions in loop quantum cosmology}

\author{Seth Connors}
\email{g_sconnors@UMassD.Edu}

\author{Gaurav Khanna}
\email{gkhanna@UMassD.Edu}
\affiliation{Physics Department, University of Massachusetts at Dartmouth, North Dartmouth, Massachusetts 02747}

\date{\today}
\begin{abstract}
In this paper we introduce a numerical approximation technique to obtain pre-classical solutions to models of loop quantum gravity. In particular, we apply the technique to vacuum Bianchi I cosmological models and recover known solutions. We also present a pre-classical solution to the Bianchi I LRS model with cosmological constant, which has not appeared elsewhere.    

\end{abstract}


\maketitle

\section{Introduction}

Loop quantum cosmology (LQC)~\cite{lqc} is a finite, symmetry reduced model of loop quantum gravity ~\cite{rov}. Study of LQC has led to the emergence of a possible mechanism for inflation~\cite{inf}, resolution of classical singularities~\cite{boj02}, as well as the development of effective semi-classical Hamiltonians~\cite{ban-dat05}. The quantum Hamiltonian constraint in LQC is a difference equation relating discrete eigenstates of the triad operators. Generic solutions to difference  equations are such that as the index parameter is increased by one, the solution will change its value drastically and possibly even change sign. This can be interpreted physically as quantum effects becoming important even at large scales in our universe. At the present time there is no physical inner product in LQC to eliminate such unphysical solutions. This has led to the notion of {\it pre-classicality}~\cite{boj01}, where solutions with this unphysical oscillatory behavior far away from the singularity are eliminated from consideration. Finding these smooth or pre-classical solutions to a discrete difference equation is a difficult problem in general, but generating function techniques~\cite{car-kha-boj04, car-kha05, dat05} provide a systematic method to obtain them. However, the method can be successfully applied to only a small subclass of models where one can obtain the generating function corresponding to the difference equation analytically. 

In this paper we present a simple numerical technique that efficiently finds approximate pre-classical solutions to any difference equation, including partial difference equations. The basic idea behind this approach is to start with a basis for the pre-classical sequence space and solve the difference equation on the linear span of this basis. The chosen basis should allow for drastic and oscillatory behavior for small values of the index parameters, but only captures smoother behavior for larger values of the parameters. One simple approach to build such a basis is to consider an ordinary fourier-time basis, but with non-constant frequency i.e. $\omega(t)$. Note that our requirement of smoother (``slower'') behavior for larger values of time $t$, then becomes the condition that $\omega(t)$ be a monotonically decreasing function of $t$. Specifically, we let $\omega=\dot\theta$ drop exponentially with time. In that case our desired fourier-time basis would have the form, $\exp(ik\theta(t))$ where $\theta(t)$ also drops exponentially with time. Given such a form for a pre-classical basis, we can then expand any  pre-classical sequence as linear combination of these basis elements and obtain the values of the coefficients by imposing the difference equation. Thus, ultimately we solve a linear system of equations which is possibly the most extensively studied problem in numeric computation with a wide variety of efficient algorithms and solvers readily available~\cite{lapack}.  

Below we apply the technique to anisotropic LQC models, in particular Bianchi I cosmology. The difference equations arising from the vacuum Bianchi I system and its locally rotationally symmetric (LRS) version~\cite{homo} can be separated into products of one-parameter equations~\cite{car-kha-boj04, car-kha05, car-kha05-2}. In addition, the Kantowski-Sachs model, which is relevant for the interior of the Schwarzschild black hole is a similar case~\cite{boj-com}. In the next section we apply our  approximation method to the one-parameter equations that arise in these separable models. In the following section we apply the method to a non-separable, two-parameter model i.e. Bianchi I LRS with cosmological constant. We end with a discussion of our results and technique.  

\section{One-parameter difference equations} 

The one-parameter equations that arise in the vacuum Bianchi I separable models are of the form~\cite{car-kha-boj04,car-kha05}, 
\begin{equation}
\label{a-eqn}
 a_\lambda (m+1) -  a_\lambda (m-1) = \lambda f(m) a_\lambda (m)
\end{equation}
where $\lambda$ is a real constant and $f(m) = 1/m$ when $m \ne 0$ and identically zero otherwise~\cite{dnfn}. Pre-classical solutions to this equation are known, both when $m$ is an integer or not~\cite{car-kha-boj04,car-kha05,car-kha05-2}. Here we will reproduce these known solutions using our approximation technique.  

We start by defining a basis for one-parameter pre-classical sequences based on the argument presented earlier. Specifically, consider the basis elements, 
\begin{equation}
\label{basis}
b_k (m) = \exp[ik\exp(-2m/M)]
\end{equation}
where $M$ corresponds to the finite length of the approximate sequence and $k$ is simply the appropriate ``fourier mode'' of the basis. The factor $2$ in the argument of the internal exponential function makes sure that over the length of the sequence, its ``frequency'' drops by an order of magnitude. Through experimentation we have observed that that particular factor is a good choice. Note that other minor variations to this form of the basis should also work correctly; there is nothing very special about the choices we made.

\begin{figure}[hbt]
	\includegraphics[width = 0.8\textwidth]{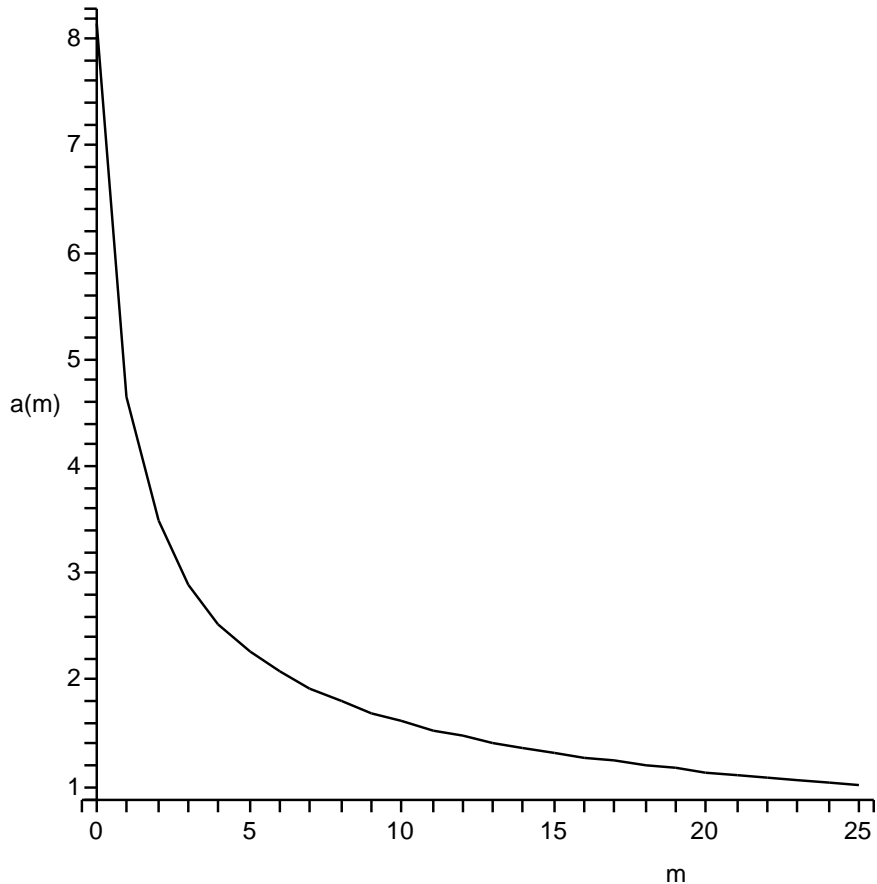}
	\caption{\label{fig1}Separable pre-classical sequence $a_{-1}(m)$ with values $\lambda = -1$, and the constraint that at large values of $m$ the sequence must approach $1$. Relative error in this solution is about one part in $10^{25}$.}
\end{figure}

Now we expand a generic solution as a complex linear sum of the basis above and substitute it into equation \ref{a-eqn}. We obtain a linear system of equations from which we can solve for the complex coefficients rather easily. Note that we still have the freedom to choose a scale for our solution and we do so by imposing an extra condition, for example $a(0)=1$. The scale-fixing conditions we use, are just a sample few of the various types of the conditions one can impose. Solution sequences obtained in this fashion are manifestly pre-classical. Below are some detailed examples.

\begin{figure}[hbt]
	\includegraphics[width = 0.8\textwidth]{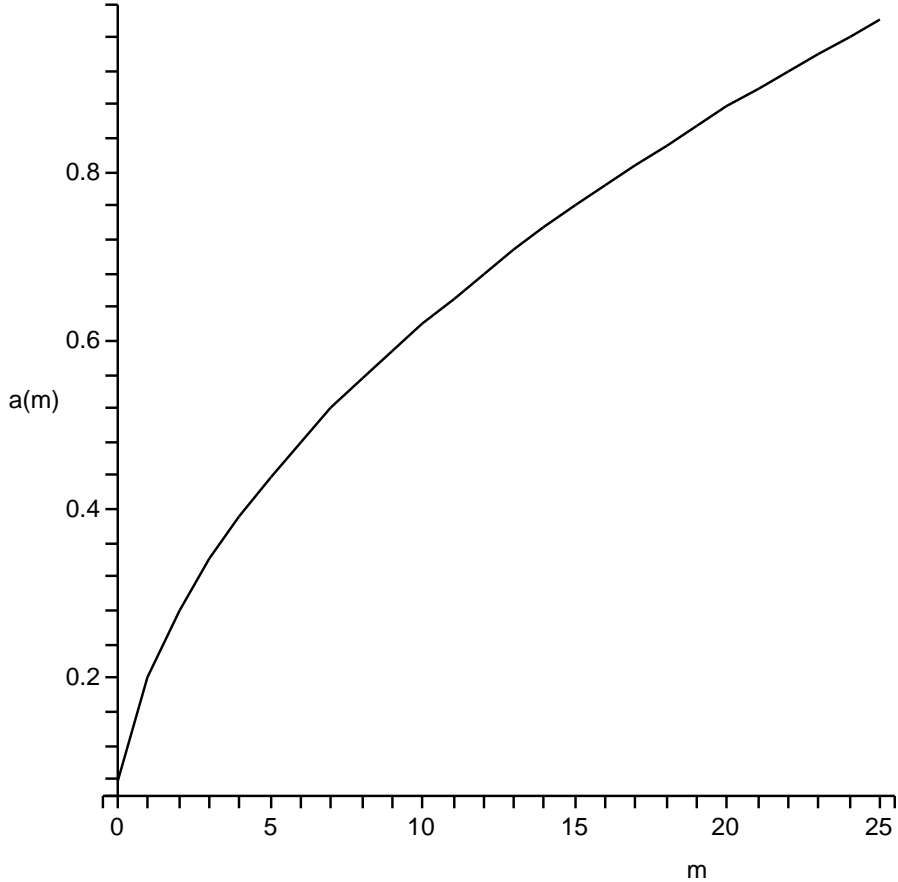}
	\caption{\label{fig2}Separable pre-classical sequence $a_{1}(m)$ with values $\lambda = 1$, and the constraint that at large values of $m$ the sequence must approach $1$. Relative error in this solution is about one part in $10^{25}$.}
\end{figure}

For figure $1$ we chose $M=25$ and imposed the scale-fixing condition $a(M)=1$ i.e. imposed that the sequence asymptotically approaches $1$. The choice of $\lambda=-1$ generically yields unphysical growing step-oscillations in the system \ref{a-eqn}, but as can be seen in the plot we obtain a smooth solution sequence. The relative error in our solution, which is of the order of $10^{-25}$ and can be made arbitrarily small solely depending on the precision of the floating point numerics chosen. For figure $2$ the only change made is in the value of $\lambda$. Again we obtain a smooth solution sequence. In figure $3$ we plot a solution from the non-integer sector of the difference equation and we solve on both sides of the origin, by extending our pre-classical basis in both regions. The scale-fixing condition here was chosen to be $a(0)=1$ while $\lambda=4$. The values of the index parameter $m=n+3/4$, where $n$ is an integer. Note that all these solutions have been obtained in the recent past using the analytic generating function approach~\cite{car-kha-boj04, car-kha05, car-kha05-2}. In that approach a very special choice has to be made for the ratio of the first two values of the sequence to obtain these smooth results. Finding this ratio takes a considerable amount of analytic computation and is often not possible in more complicated models. Using our approximate technique we obtain the desired solutions with considerable ease.
\begin{figure}[hbt]
	\includegraphics[width = 0.8\textwidth]{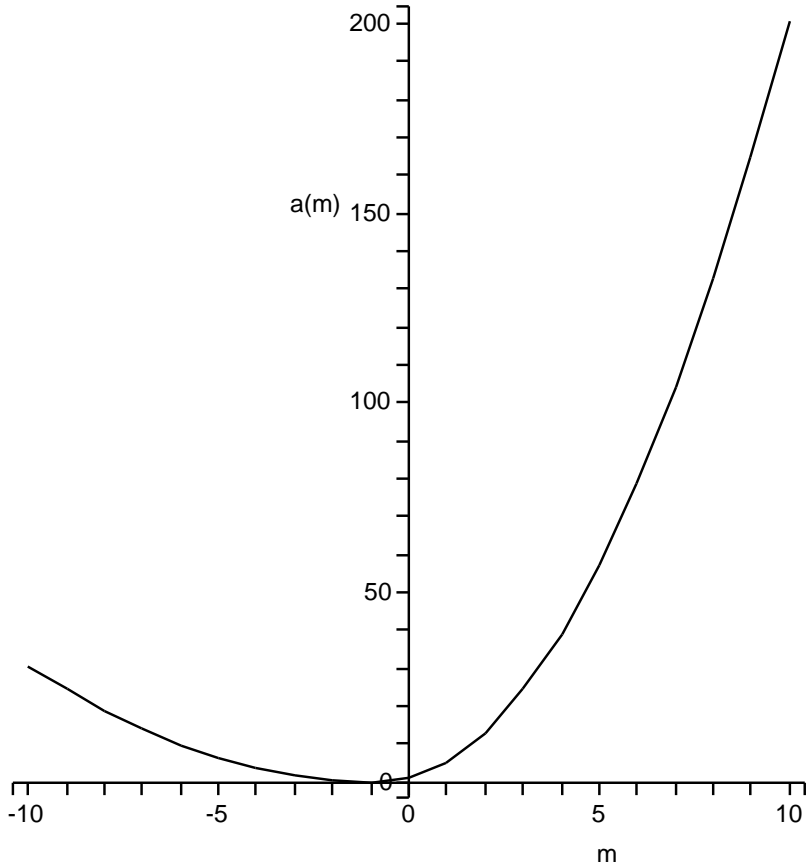}
	\caption{\label{fig3}Separable pre-classical sequence $a_{4}(m)$ with values $\lambda = 4$, $a(0) = 1$ and $m = n + 3/4$, $n$ an integer. Relative error in this solution is about one part in $10^{25}$.}
\end{figure}

\section{Two-parameter difference equations}

The two-parameter equation that arises in the Bianchi I LRS model with cosmological constant is of the form~\cite{homo}, 
\begin{equation}
  f(m) (t_{n+1,m+1}-t_{n+1,m-1})+ \frac{f(n)}{4} (t_{n,m-2}-2t_{n,m}+t_{n,m+2})\\
 + f(m) (t_{n-1,m-1}-t_{n-1,m+1}) = \beta t_{n,m}
\end{equation}
where $f(z) = 1/z$ when $z \ne 0$ and identically zero otherwise~\cite{dnfn}. Note that $\beta$ here is proportional the cosmological constant. This equation is separable if $\beta$ is zero and it reduces to the case discussed in the previous section. In this section, however, we will apply our approximation technique and find pre-classical solutions to the above equation with $\beta \ne 0$ in the integer sector.

We start again by constructing a pre-classical basis for the double-sequence space by taking products of the basis elements as shown in equation \ref{basis} in the two independent directions. We choose the 2d grid size to be $25\times25$ and $\beta=1/6$. Also the scale-fixing condition is chosen to be one of relevance for the semi-classical limit, i.e. a Gaussian wave-packet in the parameter $m$. We impose this condition at the largest value of $n$ allowed on our grid, i.e. at $n=25$. We also impose the condition that the solution vanish if $n$ or $m$ equal zero, as required by the model~\cite{homo}. 

\begin{figure}[hbt]
	\includegraphics[width = 0.8\textwidth]{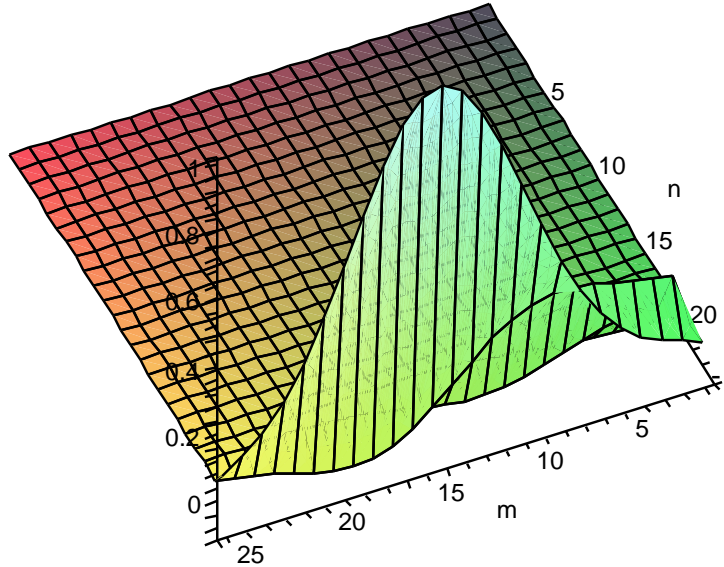}
	\caption{\label{fig4}Pre-classical double sequence solution to Bianchi I LRS model with non-zero cosmological constant. $\beta$ was chosen to be $1/6$ for this case. Relative error in this solution is about one part in $10^{8}$.}
\end{figure}

\begin{figure}[hbt]
	\includegraphics[width = 0.8\textwidth]{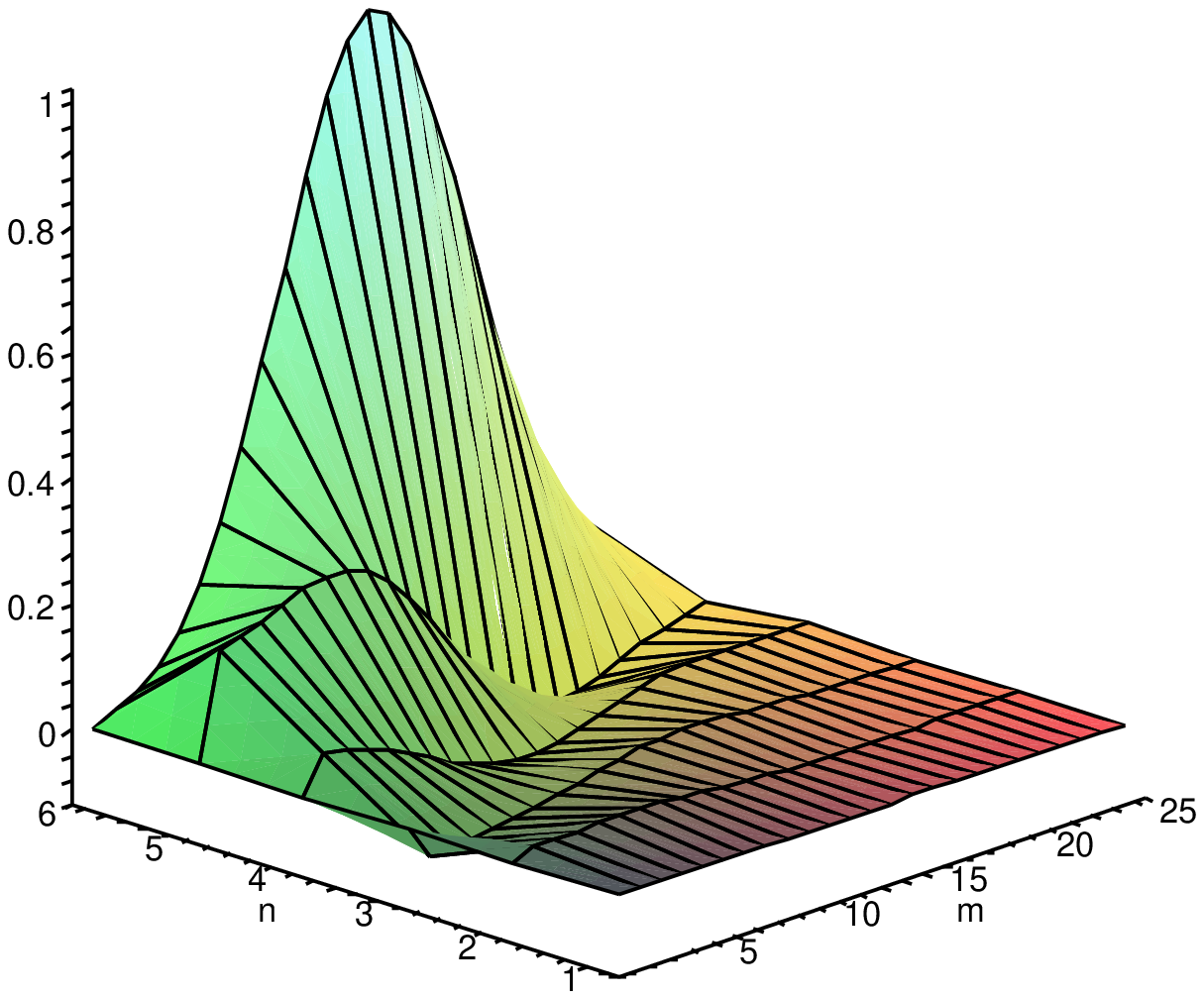}
	\caption{\label{fig5} Last six steps in $n$, from the pre-classical double sequence solution to Bianchi I LRS model with non-zero cosmological constant. $\beta$ was chosen to be $1/6$ for this case. Relative error in this solution is about one part in $10^{8}$.}
\end{figure}

The computed solution appears in figure $4$, $5$ and clearly has the desired characteristics. This is quite interesting because in the vacuum case it has been noted before~\cite{car-kha-boj04} that the integer sector lacks the variety necessary to construct arbitrary wave-packets. However, we note here that upon including a cosmological constant, that problem appears to have been alleviated. In other words, it seems plausible that addition of a cosmological constant and possibly even matter to LQC models may have the positive effect of generating a richer variety of physically meaningful solutions. Instead of presenting more plots depicting the dependence of the solution on $\beta$, we simply comment on the fact that increasing $\beta$ appears to make the solution even smoother, reinforcing our statement above.  

Another approximate way to study the effects of the inclusion of a cosmological constant, is to consider the additional term as a perturbative term in the Wheeler-DeWitt equation of this model. Without the cosmological constant term one obtains only certain restrictive forms for the $m$-dependence of the solution, i.e. only power-law decays~\cite{car-kha-boj04}. However, if the effect of including a small cosmological constant term is studied perturbatively, one notes the emergence of a richer variety in the $m$-dependence of the solution. In particular, positive powers of $m$ appear in addition to the original negative ones, suggesting the  possibility of the existence of a larger solution space. 

Note that extending the pre-classical basis on both sides of the origin can be done in the non-integer sector of this two-parameter model as well. However, that imposes the condition that the solution must be pre-classical on both sides of the classical singularity. This may be overly restrictive. The question of whether solutions to cosmological models should be pre-classical on both sides of the classical singularity or not, is open for discussion. If one feels that in the distant ``past'' before the big bang, physics at scales comparable to the current scale of the universe would necessarily be describable by partial differential equations then pre-classicality on both sides of the singularity would be a physical requirement. In that case, the vacuum homogeneous models we have discussed here appear to lack the variety that one would desire its pre-classical solutions to have~\cite{car-kha05-2}. On the other hand, if one insists on pre-classicality only on one side of the singularity, then there is no such issue if one includes the solutions from the non-integer sector of the theory. In that case, we simply note the physical significance of the non-integer sector of LQC~\cite{dat05,car-kha05-2}. 

\section{Discussion}

We have demonstrated an efficient, numerical approximation technique to compute pre-classical solutions to generic LQG models. The method is able to reproduce known solutions and generate new ones, to an arbitrary degree of accuracy. In this method one ultimately solves a linear system numerically, which is a very well researched problem in computation, with several well established and highly efficient algorithms and software packages readily available~\cite{lapack}. 

We also observed that the inclusion of a cosmological constant in an anisotropic model of LQC i.e. Bianchi I LRS with the cosmological constant appears to yield a richer class of pre-classical solutions compared with the vacuum case. It is possible that inclusion of matter in LQG models has the desirable effect of enlarging the physically meaningful solution space. 

\section{Acknowledgments}

The authors appreciate the helpful comments from Daniel Cartin. The authors are grateful for research support from the University of Massachusetts at Dartmouth, as well as Glaser Trust.


\begin{thebibliography}{99}
\bibitem{dnfn} Note that the function $f(n)$ here is an approximation to the function $d(n)$ as used in reference~\cite{homo}. However, this makes no significant difference for our purposes.
\end{thebibliography}
\end{document}